# Energy exchange between surface plasmon polaritons and CdSe/ZnS quantum dots

Kunal Tiwari[1,2], Suresh C Sharma[1,*], Hussein Akafzade[1], and Nader Hozhabri[3]

[1]Department of Physics, University of Texas at Arlington, Arlington Texas 76019 USA

[2]Wireline Research & Development, Weatherford, Int., Katy, Texas 77493 USA

[3]Nanotechnology Research Center, Shimadzu Institute, University of Texas at Arlington, Arlington Texas 76019 USA

*Author to whom correspondence should be addressed at sharma@uta.edu

## ABSTRACT

Evidence is presented for energy exchange between surface plasmon polaritons (SPPs) excited in waveguide-coupled $Ag/Si_3N_4/Au$ structures and CdSe/ZnS quantum dot (QD) emitters. The QD dispersion is coated on the surface of $Ag/Si_3N_4/Au$ nanostructure, surface plasmon resonance (SPR) is excited on surface of multilayered structure and the photoluminescence (PL) emission by the QDs is monitored as a function of the evanescent electric field by using fixed-detector pump-probe spectroscopy. We observe relatively large shift in the PL emission wavelength (corresponding to 4-5 meV) with accompanying electric field induced quenching (up to 50%) of the emission. Under the influence of the evanescent field, the PL emission peak splits into two components corresponding to two frequencies (Rabi splitting). Computer simulations are used to understand conditions under which Rabi splitting should occur.

*Corresponding author (sharma@uta.edu)

1. Introduction

The performance of photovoltaic devices and light-emitting diodes (LEDs) can be improved by efficient energy transfer between surface plasmon polaritons (SPPs) and semiconducting quantum dots (QDs). The surface plasmon polaritons are collective oscillations of the surface charges, which propagate along an appropriate metal/dielectric interface. These excitations have associated with them evanescent fields that decay exponentially along the normal to the interface.[1-5] It is well known that the fluorescence emission by a molecule can be greatly influenced by the electromagnetic fields in its surroundings.[6, 7] One of the important questions in this context relates to possible modifications in light emission, photoluminescence (PL) by the QDs because of coupling between the evanescent fields of the SPPs and the emitters.[8-10] It is known that the strength of the coupling between the emitters and the evanescent fields can be increased by: (a) confining light into smaller volumes and/or (b) increasing the dipole moment (oscillator strength) of the emitters. Whenever a sufficiently large number of emitters is "embedded" within strong electromagnetic fields of SPPs, these conditions are readily met. In such case, the light is confined within a small volume and because of the usage of relatively high concentrations of the QDs, the net dipole moment of the ensemble of the emitters is increased. Therefore, the evanescent electromagnetic fields of the SPPs are expected to produce significant changes in the PL spectra of the QDs. The results presented in this publication show clear evidence for the resonance energy exchange between the electromagnetic fields of the SPPs and CdSe/ZnS emitters. We observe large redshifts in the PL emission wavelength and strong quenching of the PL intensity under experimental conditions of resonance excitation of the SPPs. Our results confirm that the light emission by CdSe/ZnS QDs can be tuned remarkably well by resonant energy exchange between the evanescent fields and the QDS, especially when certain waveguide-coupled multilayer structures are used to excite SPPs in proximity of the emitters. These data further show that the PL tuning over wide range of wavelengths is accomplished simply by scanning the angle at which a *p*-polarized laser strikes the multilayer structure in the Kretschmann experimental setup. The motivation for this work comes from: (a) as mentioned above, a quest for an efficient and tunable energy exchange mechanism between light emitters and evanescent fields, (b) observation of an order-of-magnitude higher red-shift in the central wavelength of the photoluminescence spectra of CdSe/ZnS QDs coated on the surface of waveguide-coupled multilayer structure (present work), and (c) desire to better understand nature of the coupling between CdSe/ZnS emitters and evanescent fields. The CdSe/ZnS QDS are potential candidates for blue-green semiconductor diode lasers and LEDs. In the core-shell structure of the CdSe nanocrystals (bandgap = 1.7 eV @ 300K), capped with a thin layer of larger bandgap semiconductor, ZnS (bandgap = 3.68 eV @ 300K)[7], whereas the holes are confined within the CdSe core due to passivation of the QD-surface by ZnS layer, the electron wavefunction extends well into the ZnS shell. [11]This extension depends on several parameters, e. g., shell material (ZnS *vs* CdS), band offsets in the core-shell structure, and dielectric constant of the QDs *etc*. The outermost TOPO layer passivates the outermost CdSe/ZnS surface states and further enhances PL emission and stability of the nanostructure. In an earlier experiment, we had studied PL from CdSe/ZnS QDs by using pump-probe laser beams and fixed detector Kretschmann optical system using single-metal Au/high-index prism SPR sensor.[12] These experiments provided two important results: (1) PL emission by CdSe/ZnS QDs is red-shifted by relatively small but significant amount (approximately 0.1 nm Stark effect in the 1s-1s transition energy in the HOMO-LUMO molecular orbital energy of the QDs) and (2) PL intensity is quenched almost exponentially with angle of incidence as it is varied away from $\theta_{ATR}$, the angle at which total internal reflection occurs. Based on reasonable arguments, we estimated that the observed red shifts and quenching would require evanescent fields of approximately 20 kV/cm. Others have obtained similar results and they are reasonably well understood.[8, 9, 13, 14] For certain other materials, just the opposite effect is observed; PL intensity increases and the emission wavelength is shortened (blue shifted). Whether the PL intensity is enhanced or quenched is determined by the balance between the resonant energy transfer from the QDs to the metal, as well as by the enhancement in the electric field associated with SPPs.[15] In the case of the *so-called* type-I systems, which include CdSe/ZnS quantum dots, the quantum-confined Stark effect causes a quadratic redshift in the PL emission.[16] On the other hand, in type-II systems, which include CdSe and PbS quantum dots, the electric field causes a blue shift.[17, 18]

2. Experimental Details

In earlier work, we have utilized computer simulations and optical SPR measurements to compare SPR characteristics of several different types of materials and structure, such as single metal (Ag or Au), bi-metal (Ag/Au) and waveguide-coupled $Ag/Si_3N_4/Au$ multilayer structures.[19, 20] In the context of the present work, we discovered that the strength and decay length of the evanescent fields can be increased by several times (compared to the fields generated at the dielectric/single-metal surfaces) by using optimized thickness of multilayer structures, like the 35nm Ag/150nm $Si_3N_4$ dielectric/28nm. In these structures, the evanescent fields at the surface of the sensor are 3-5 times stronger and decay exponentially with distance from and normal to the sensor surface with decay lengths $\geq$ 300 nm.

The quartz/Ag/$Si_3N_4$/Au multilayer waveguide-coupled structures were fabricated in *Class-100 NanoFab* and coupled to high-index prism for optical measurements. Details on the growth, characterization, and SPR measurements on several different sensors fabricated in *class-100* clean room have been presented elsewhere.[21-23] The QDs' dispersion in Chloroform was dropped on the sensor surface forming a surface coating. Special attention was paid to deposit, as thin as possible, film to minimize interparticle interactions. The PL and SPR measurements were carried out by using two experimental set ups; (1) a fixed detector optical system (shown in figure-1), which enables simultaneous measurements of SPR and PL spectra without the requirement for a $(\theta, 2\theta)$ goniometer.[12] This system has the added advantage that in its pump-probe configuration, both SPR and PL spectra are excited at the same spot on the surface of the sensor (to within $\leq 0.5$ $\mu$m), and (2) a slightly modified pump-probe system to apply external *ac* electric fields and measure PL spectra as functions of the strength of the applied field. This system is shown in figure 2. We used two different interrogation wavelengths, $\lambda = 514$ and 632 nm in the depicted pump-probe configuration. The QDs are pumped by the 514 nm radiation from an argon-ion laser. The SPPs are excited on the surface of the sensor by using *p*-polarized 632 nm radiation from He-Ne laser. The coupling between the PL emission and the SPPs is studied by measuring simultaneously PL spectra and SPR curves as functions of the angle of incidence for *p*-polarized excitation laser in the Kretschmann configuration optical system sketched in figure 1.

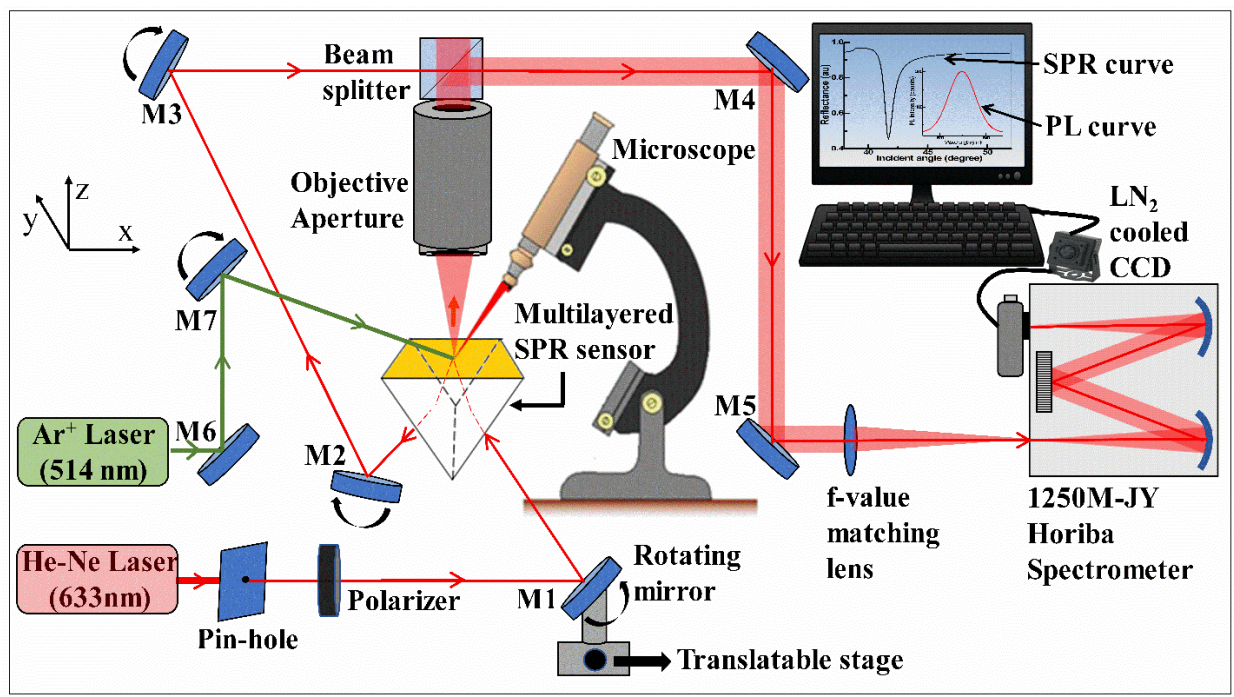

Figure-1. Pump-probe optical spectrometer based on the fixed-detector SPR measurement Kretschmann configuration system

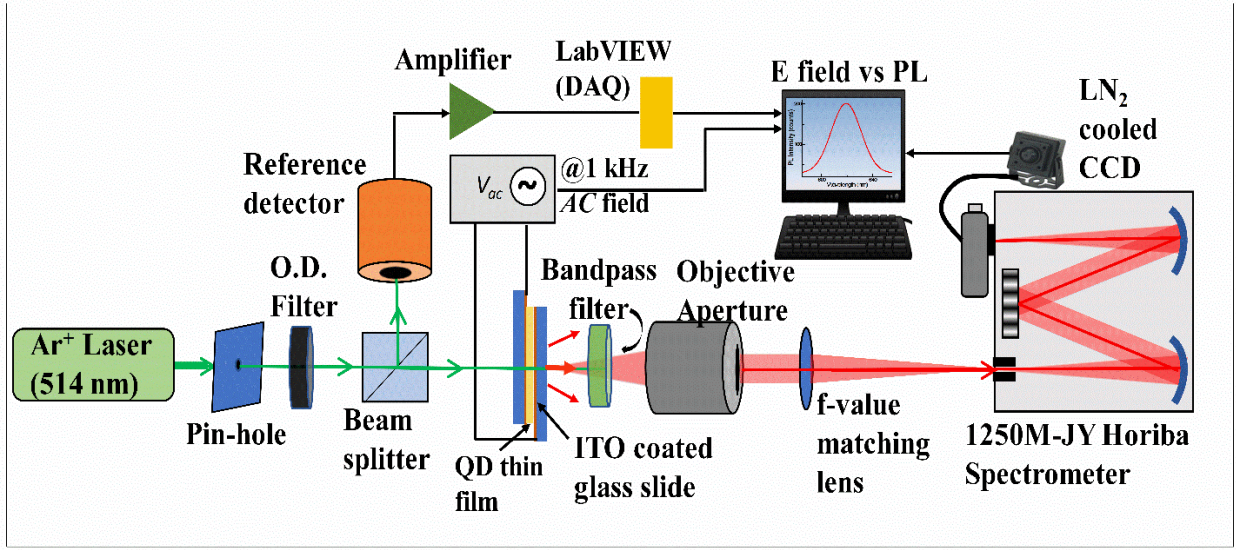

## 3. Results and discussion

### 3.1.1. Microstructure and absorption

Figures 3-4 show results of SEM and EDS analyses of a thin film, containing QDs, deposited over the $Ag/Si_3N_4/Au$ sensor structure. It is evident from these results that the film is indeed thin. The SEM picture shows small clusters in the film. The EDS analysis shows relatively weak signatures of CdSe/ZnS, but a strong signal from the underlying elements of the multilayer nanostructure of the sensor ($Ag/Si_3N_4/Au$) upon which the film was deposited. For a thick film, obviously, the signal from these elements of the sensor nanostructure will not be visible. The absorption spectrum

of the QDs in the 400–700 nm range is shown in figure 5. There is a significant absorption at both exciting wavelengths, 514 and 632 nm.

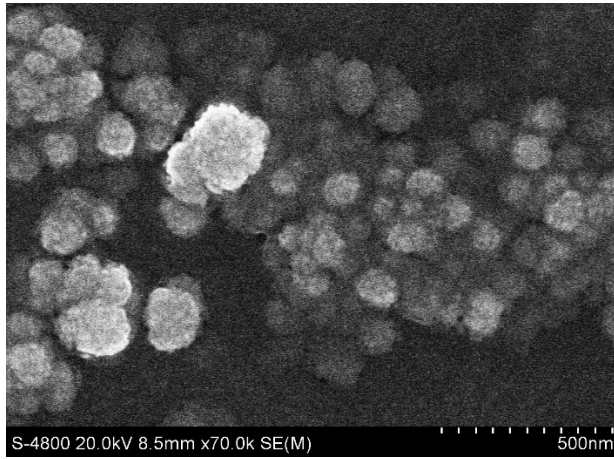 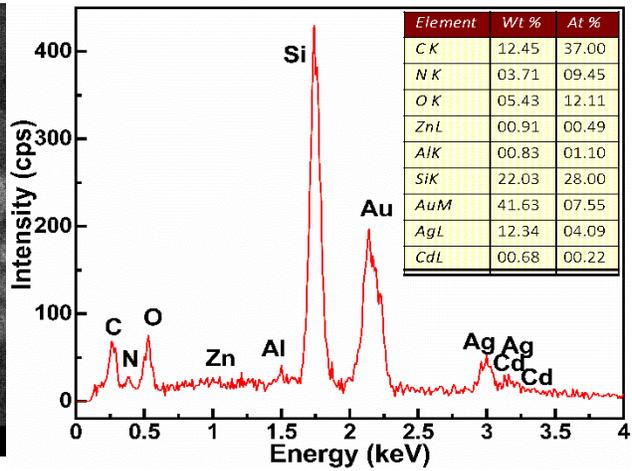

Figure 3. SEM picture of CdSe/ZnS QDs deposited over Ag/Si$_3$N$_4$/Au nanostructure

Figure 4. EDS analysis of QDs thin film coated on Ag/Si3N3/Au sensor structure

3.1.2. PL spectrum

Figure 6 shows the PL spectrum for an excitation wavelength of 514 nm and emission peak occurring at 618.5 nm. The PL spectrum, in the presence of the surface plasmon resonance, is shown in figure 7 (a)-(b). The prism and nanostructure sensor schematically show that the SPR is excited by *p*-polarized 632 nm and the 514 nm radiation is absorbed by QDs leading to PL emission at approximately 620 nm. The inset shows PL emission peaks in the absence and presence of the surface plasmon resonance.

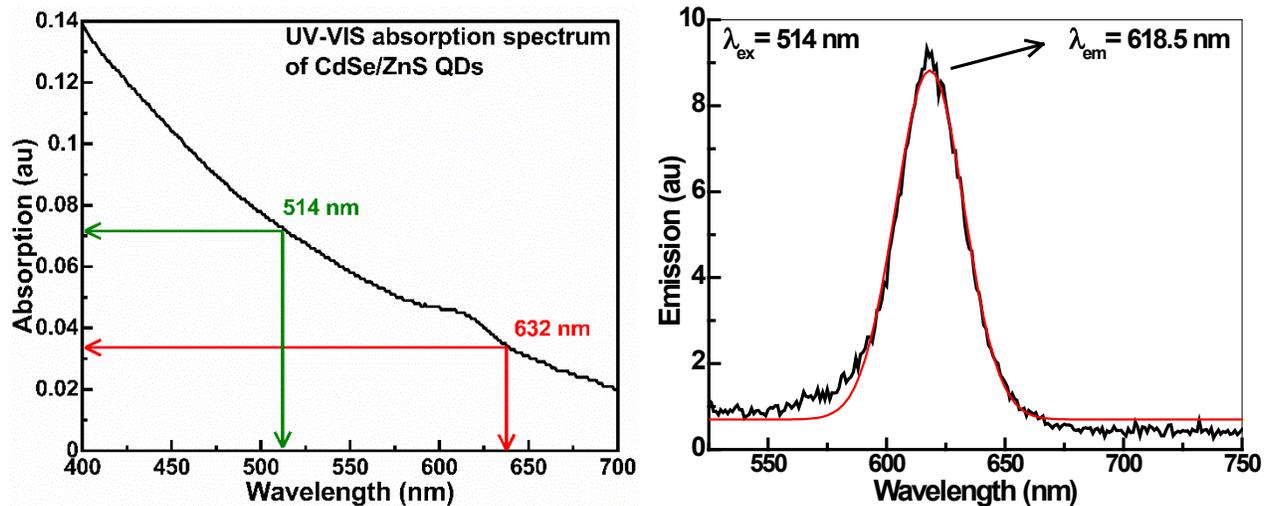

Figure 6. PL spectrum of QDs pumped by 514 nm green laser.

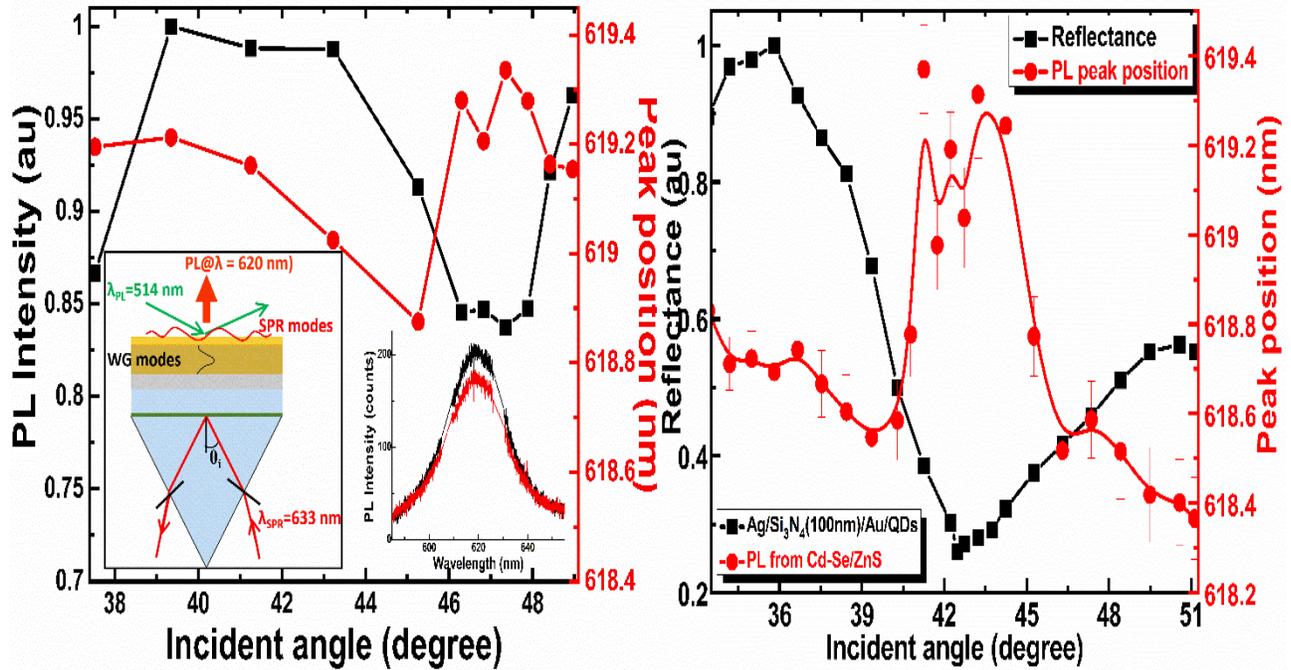

Figure 7 (a). Schematic of the prism setup to create SPR and PL emission, The PL spectra show SPR-induced quenching.

Figure 7 (b). The peak wavelength of the PL spectrum (circles) and SPR as a function of the angle of incidence.

The maximum quenching of the PL emission is observed at resonance $\theta = \theta_{SPR}$, where $\theta_{SPR}$ is the resonance angle, approximately $47^0$ in figure 7 (a). Again as shown in figure 7 (a), the PL emission wavelength is blue shifted for incident angles between $38^0$ and $45^0$. For angles between $45^0$ and $48^0$, the emission wavelength is red shifted. The data of figure 7 (b) show the SPR and PL emission wavelength as a function of the incidence angle for Ag/Si$_3$N$_4$/Au nanostructure coated over the prism. Again, the emission wavelength is blue shifted for angles between about $34^0$ and $40^0$. For angles between approximately 40 - $47^0$, the emission wavelength is red shifted from about 618.4 nm to 619.4 nm.

### 3.1.3. Measurements of PL spectrum as function of externally applied electric field

In order to better understand the strength of the evanescent fields required to produce the observed shift in PL spectrum, we conducted an experiment in which PL spectra were measured as function of externally applied electric fields. Figure 2 shows the block diagram of the experimental system used to carry out these measurements. A thin film containing QDs is deposited over the conducting surface of an Indium-Tin-Oxide (ITO) coated glass slide. This is covered by another ITO coated glass slide so as to have QDs inside a parallel plate capacitor, across which *ac* electric field (1 kHz) is applied. The PL emission wavelength is shown as a function of applied electric field in figure 8. The results show that an electric field of approximately $10^6$ V/m is required to induce a blue shift of about 0.6 nm. The results are reminiscent of the Rabi splitting in the dispersion curves. We have employed COMSOL software to simulate the dispersion curves for the QDs and access the condition for Rabi splitting in these curves.[6, 24-28] In the following, we present details of the simulations and resulting data.

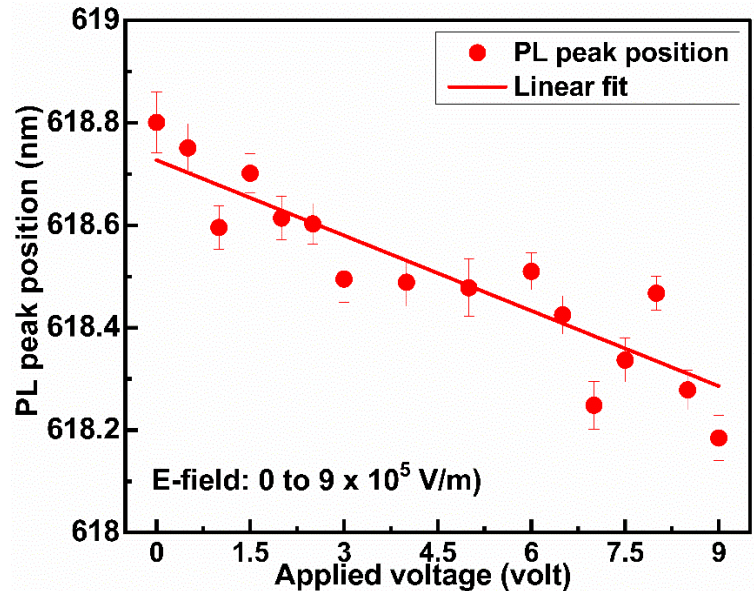

Figure 8. PL peak wavelength as a function of applied electric field

## 4. Rabi splitting in the dispersion curves

The simulations for Rabi Splitting were carried out for QDs probed by using the Kretschmann configuration with a high index prism of refractive index = 1.7847 covered by 48 nm gold layer and air as the ambient dielectric material. An array of emitters was positioned with 10 nm interparticle distance on the surface of the gold film. The QDs are assumed to have the freedom to oscillate under the influence of the evanescent electric field created by surface plasmons. The classical Lorentz-Drude model offers a good approximation to qualitatively understand light matter interactions.[29-34] In the case of the surface plasmon polaritons, this model considers them as an array of oscillators, each of mass m, oscillating with a natural frequency $\omega_0$ under the influence of a restoring electric force with frequency $\omega$ and coupling constant $\gamma$. The equation of motion can be represented by,

$$m\ddot{r} + m\gamma\dot{r} + m\omega_0^2 r + eE(r,t) = 0$$

Solving this differential equation leads to an expression of dipole moment, given by,

$$P = -er = -\frac{e^2}{m}\frac{E_0 e^{-i\omega t}}{\omega_0^2 - \omega^2 + i\gamma\omega},$$

in which $e$ and $m$ are the magnitude of the charge and mass of electron respectively. The frequency of the incidence light $is\ \omega$. The evanescent electric field $E_0$ decays exponentially with distance from the sensor surface along the z-direction; for a prism of 1.7847 index, coated with 48 $nm$ gold film and incident laser with $\lambda = 632.8 nm$, the evanescent field decays with a decay length of approximately 400 $nm$. There is also an electric field in the plane of the film, along the x-direction, but it's much smaller in magnitude. The following figures show representative results simulated for 10 mW laser with a spot size of 9 mm² cross sectional area incident upon the sensor.

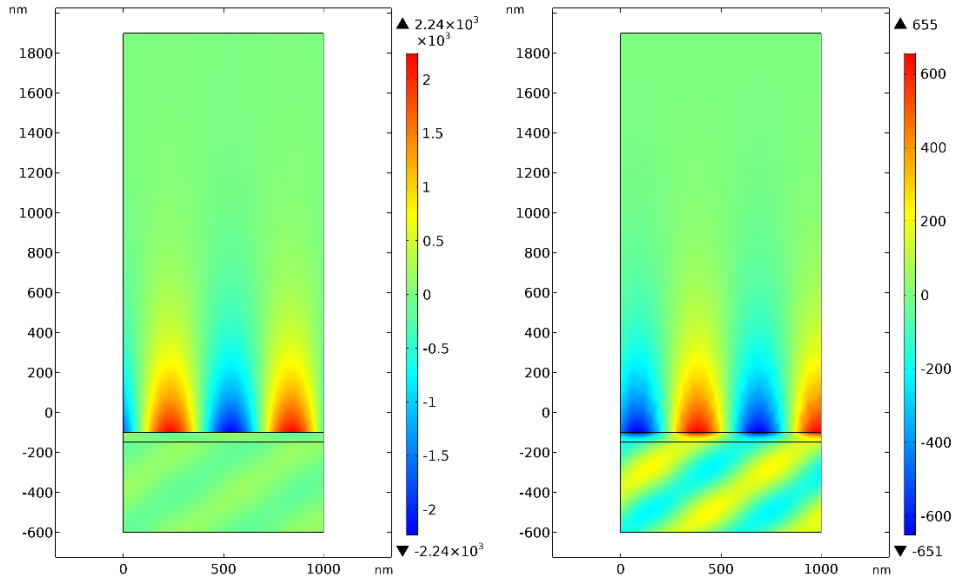

Figure 9. The evanescent electric field along the z-direction

The same logic used for surface plasmon oscillations can also be used to describe the single electron oscillations inside a QD. The electrons are treated as emitters coupled to the evanescent electric filed $E$ and their dipole moment is given by,

$$P = -\frac{e^2}{m}\frac{E}{\omega_0^2 - \omega^2 + i\gamma\omega},$$

in which $\omega_0$ is the frequency of PL emission and $\gamma$ is the damping constant for the quantum dot. Since quantum dots have usually the size of 10 nm or smaller, the magnitude of the electric field that they experience can be assumed to a good approximation as constant on the surface of the gold film. Simulation results for maximum strength of the evanescent waves in the z-direction on 48 nm gold film for incidence wavelength range of 550 nm to 750 nm and angle of incidence between 35 and 37 degree are shown in figure 10. The real part of the gold dielectric constant does not have high enough magnitude to show strong SPR effect for the wavelengths less than 550nm.

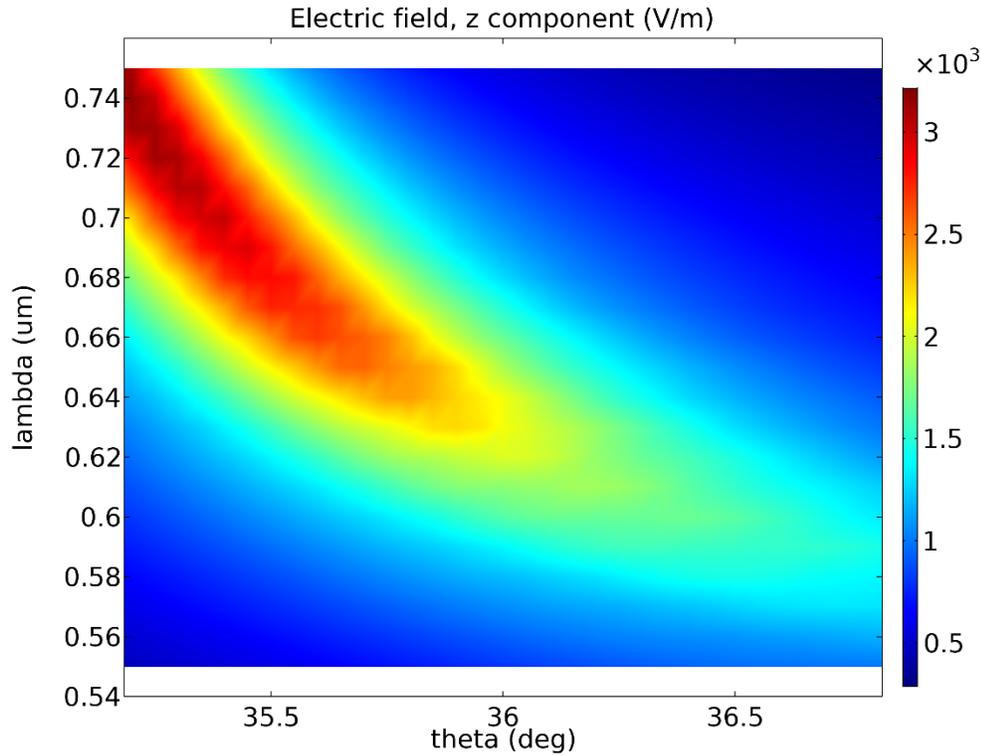

Figure 10. Maximum E-field magnitude on the surface of the 48 nm gold film

We use the following two independent variable function to describe this plot,

$$E_{z,max} = 10^3 \times e^{-\frac{(\theta-35.2\times(0.75-\lambda))^2}{(1+10\times(0.75-\lambda))^2}} \times [44.026\left(\frac{1}{\lambda}\right)^3 - 210.24\left(\frac{1}{\lambda}\right)^2 + 327.26\left(\frac{1}{\lambda}\right) - 163.63)]$$

Where the laser wavelength $\lambda$ is in $\mu m$, $\theta$ is in degrees, and the field itself is in $V/m$ units. From our measurements discussed above, the emission wavelength for the CdSe/ZnS is 620 nm and the damping constant is assumed to be on the order of $10^{13}$Hz. Figure 11 below shows the meshed unit block of our structure. The top layer is the 1.7847 index prism. A gold film with 48 nm thickness is located underneath the prism and it is in contact with air, which includes a thick layer at the bottom of the unit. The Floquet periodic condition is applied to vertical sides of this block, while the incidence light is approaching from the top side with some angle with respect to the normal, and the transmitted light is collected from the bottom side.

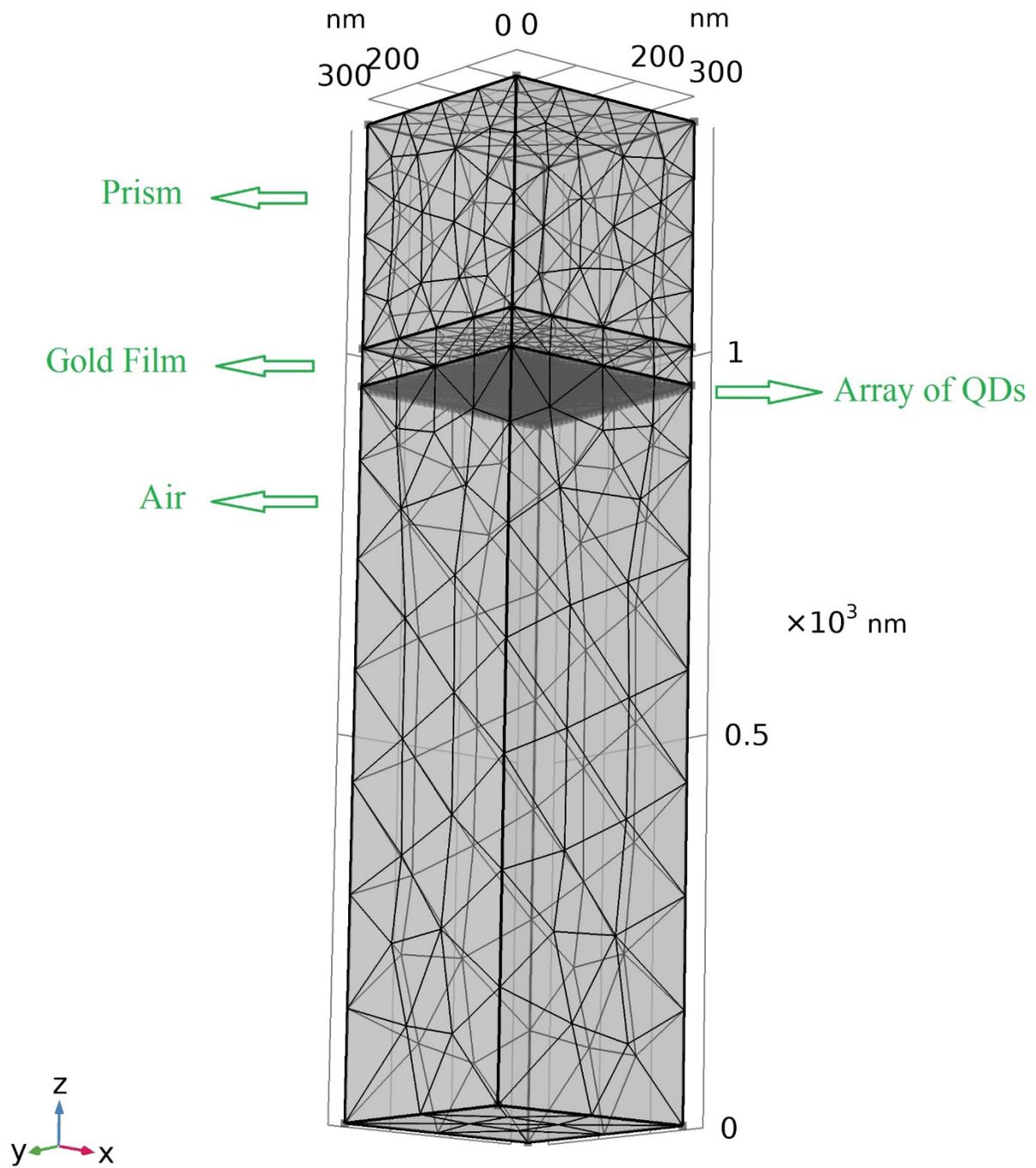

Figure 11. Meshed structure of SPPs-QDs coupling

The emitters are located within 10 nm proximity of the gold film with interparticle separation of 10 nm. The coupling strength here between SPPs electric field and QDs shows up in the number of electrons inside QD, which are engaged in oscillations and emission. A factor $N$ is inserted in front of electric current dipole moment to represent this coupling:

$$ECDM = N\frac{e^2}{m}\frac{Ef}{\omega_0^2 - \omega^2 + i\gamma\omega},$$

For the case of $N = 0$ (no QDs), the simulation is run for reflectance *vs.* angle of incidence and it produces results shown in figure 12 (a). The next sections of the figure correspond to $N = 10, 20,$ and $30$.

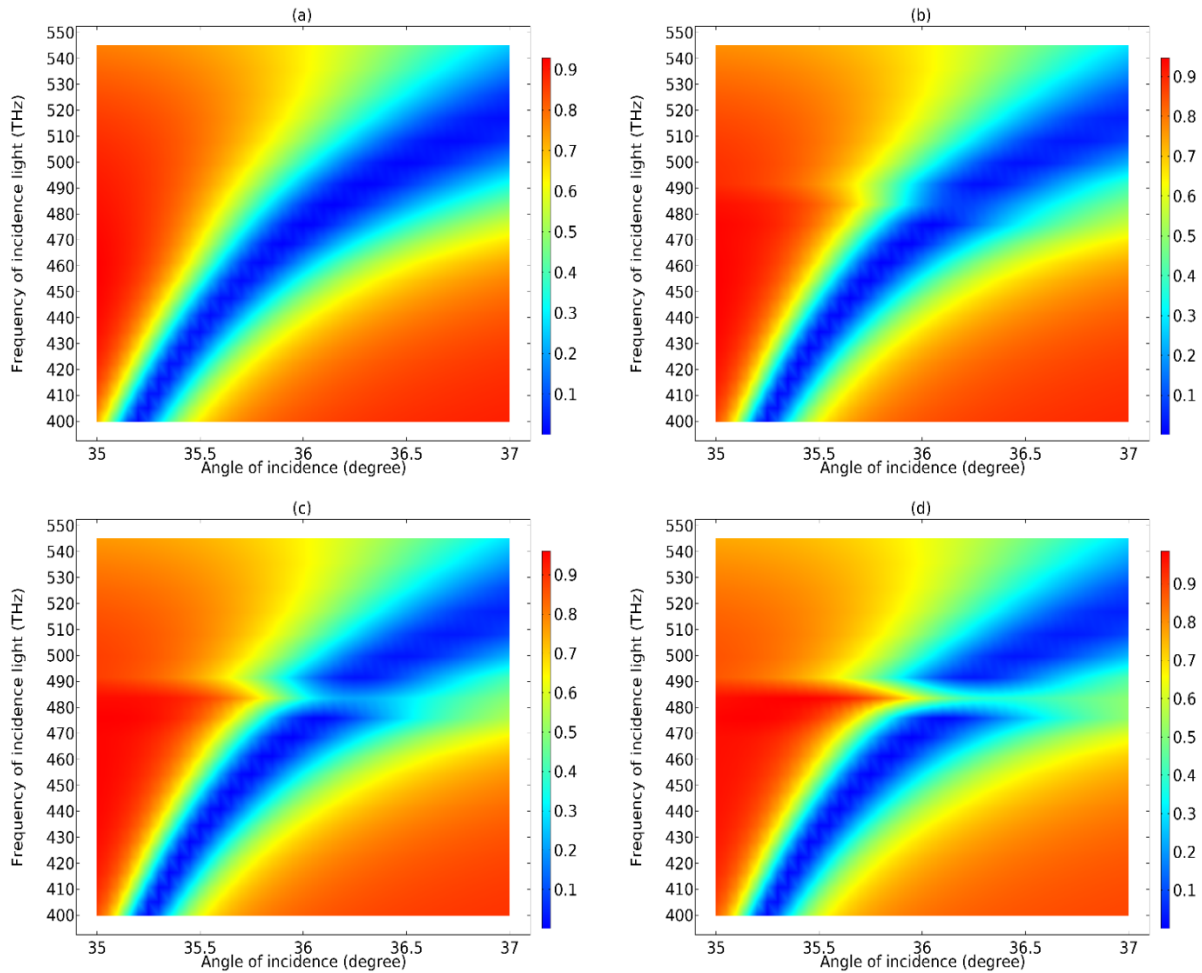

As the number of the electrons involved in the coupling increases, we see a more pronounced splitting in the dispersion curves (higher magnitude of the Rabi splitting). In part (d) of the figure, we see a splitting of about 10 THz, which corresponds to energy shift of about 40 meV. Working with single wavelength of 632.8nm and creating surface plasmons with the same 48 nm thick gold film, we ran another simulation to investigate SPPs-QD coupling. Covering the surface of the gold film with the same quantum dots and the same SPR conditions. For the no coupling case, the CdSe/ZnS quantum dots absorb a broadband light spectrum and emit mainly red light with a gaussian distribution centered at 620 nm. We observe that the single gaussian PL distribution splits into two closely centered gaussians. [16] Additionally, the PL is slightly quenched. The shift in the central wavelengths of these two gaussians can be justified by the electric field induced stark effect, given by[35],

$$\text{Energy Shift} = -\frac{9}{4}\varepsilon_2 a_{ex}^3 E^2,$$

The shift in the PL wavelength is a quadratic function of the electric field. Here $\varepsilon_2$ and $a_{ex}$ depend on the type of the QD. By using reasonable values of these parameters, the PL shift is on the order of 100 meV.[36] The shift observed in our data is much smaller, only about a few meV. Besides the electric field induced stark effect, another plausible reason for the splitting of the PL spectrum into two gaussians can be the fact that the evanescent field changes direction over a distance on the order of 250 nm. Consequently, it is conceivable that the quantum dots that happen to be under the influence of the field pointing downward behave differently compared to the adjacent QDs being acted upon by upward pointing field. Furthermore, there are published results showing that an external electric field of order $10^7$ V/m is required to produce PL shifts on the order of 2-3 meV, which is about the shift seen in our data. But our simulations show orders of magnitude lower electric fields; only about $10^3$ V/m. The reasons behind this discrepancy are not known. It is possible that the assumption made in the simulations that the surface of the gold film is smooth on nm scale is not a good approximation. The actual gold film deposited in our experiments is probably smooth only on several nm scale. Therefore, the actual gold film used has numerous rough areas with much higher local evanescent electric fields resulting varying values of shift in the PL spectrum.

## 5. Conclusions

The photoluminescence emission characteristics of CdSe/ZnS quantum dot emitters have been studied by utilizing pump-probe spectroscopy. When pumped by 514 nm green laser, the QDs emission occurs at 620 nm. In the presence of the surface plasmon resonance, the PL emission is quenched, and the emission wavelength shifts. When acted upon by the surface plasmon evanescent electric field of certain magnitude, the PL spectrum exhibits Rabi splitting. The conditions for the Rabi splitting to occur have been investigated through computer simulations of the dispersion curves.

20. Tiwari, K., S.C. Sharma, and N. Hozhabri, *Hafnium dioxide as a dielectric for highly-sensitive waveguide-coupled surface plasmon resonance sensors.* AIP Advances, 2016. **6**(4): p. 045217-.
21. Tiwari, K., *Bimetallic waveguide-coupled sensors for tunable plasmonic devices*, in *Physics*. 2015, University of Texas at Arlington.
22. Tiwari, K., S.C. Sharma, and N. Hozhabri, *High performance surface plasmon sensors: Simulations and measurements.* Journal of Applied Physics, 2015. **118**(9).
23. Tiwari, K., S.C. Sharma, and N. Hozhabri, *Hafnium dioxide as a dielectric for highly-sensitive waveguide-coupled surface plasmon resonance sensors.* Aip Advances, 2016. **6**(4).
24. www.comsol.com, *COMSOL Multiphysics*. 2019. p. A general-purpose simulation software for modeling designs, devices, and processes in all fields of engineering, manufacturing, and scientific research.
25. Hussein Akafzade and S.C. Sharma, *New Metamaterial as a Broadband Absorber of Sunlight with Extremely High Absorption Efficiency.* AIP Advances, 2020. **10**: p. 035209.
26. Sharma, S.C., H. Akafzade, and N. Hozhabri, *Waveguide-coupled Ag/HfO$_2$/Au tapered nanostructures for high-resolution surface plasmon resonance sensor applications*, in *Optical Society of America: Sesnors and Sensing Congress*. 2019, Optical Society of America: San Jose, California.
27. Sharma, S.C., et al. *Nanostructures for highly sensitive surface plasmon resonance sensors and confinement of IR energy in twodimensional materials*. in *Latin America Optics and Photonics Conference, https://www.osapublishing.org/abstract.cfm?uri=LAOP-2018-W3B.3*. 2018. Lima: Optical Society of America.
28. Akafzade, H., *Investigations of the compression of IR energy onto surface plasmons on graphene*, in *Physics*. 2018, The University of Texas at Arlington.
29. Abrikosov, A.A., *Introduction to the theory of normal metals*. Solid state physics Supplement. 1972, New York,: Academic Press. xi, 293 p.
30. Ashcroft, N.W. and N.D. Mermin, *Solid state physics*. 1976, New York,: Holt. xxi, 826 p.
31. Harrison, W.A., *Elementary electronic structure*. 1999, Singapore ; River Edge, NJ: World Scientific. xx, 817 p.
32. Bendow, B. and B. Lengeler, *Electronic structure of noble metals and polariton-mediated light scattering*. Springer tracts in modern physics. 1978, Berlin ; New York: Springer-Verlag. vi, 114 p.
33. Ziman, J.M., N.F. Mott, and P.B. Hirsch, *The Physics of metals*. 1969, London,: Cambridge U.P.
34. Hecht, E., *Optics*. 5 ed. 2017, Boston: Pearson Education, Inc. vi, 714 pages.
35. Empedocles, S.A. and M.G. Bawendi, *Quantum-confined stark effect in single CdSe nanocrystallite quantum dots.* Science, 1997. **278**(5346): p. 2114-2117.
36. Pokutnyi, S.I., et al., *Stark effect in semiconductor quantum dots.* Journal of Applied Physics, 2004. **96**(2): p. 1115-1119.
20